\documentclass[a4paper,12pt]{article}

\usepackage{url,amssymb,amsfonts,amsmath,epsfig,float,graphicx}
\usepackage[margin=0.8in]{geometry}

\newcommand{\be}{\begin{eqnarray}}
\newcommand{\ee}{\end{eqnarray}}
\newcommand{\Tr}{\,{\rm Tr}\,}
\newcommand{\upd}{{\rm d}}

\def\Co{{\mathbb C}}
\def\Eo{{\mathbb E}}

\def\Io{{\mathbb I}}

\def\Mo{{\mathbb M}}
\def\No{{\mathbb N}}
\def\Qo{{\mathbb Q}}
\def\Ro{{\mathbb R}}
\def\Xo{{\mathbb X}}

\newtheorem{theorem}{Theorem}{}
\newtheorem{proposition}{Proposition}{}
{}
{}
\newtheorem{definition}{Definition}{}
\newtheorem{example}{Example}{}

\def\beginproof{\par\strut\vskip 0.01cm\noindent{\bf Proof}\par\noindent}
\def\endproof{\par\strut\hfill$\square$\par\vskip 0.1cm}
\newcommand{\address}[1]{}
\newcommand{\keywords}[1]{}

\begin{document}

%%%%%%%%%%%%%%%%%%%%%%%%%%%%%%%%%%%%%%%%%%%%%%%%%%%%%%%%%%%%%%%%%%%%
\title{DATA SET MODELS AND EXPONENTIAL FAMILIES IN STATISTICAL PHYSICS AND BEYOND}

\author{\footnotesize Jan Naudts and Ben Anthonis\\
\footnotesize University of Antwerp, Physics Department\\
\footnotesize Universiteitsplein 1, 2610 Wilrijk-Antwerpen, Belgium\\
\footnotesize jan.naudts@ua.ac.be, ben.anthonis@ua.ac.be}
\date{}

\maketitle

\begin{abstract}
The exponential family of models is defined in a general setting, not relying on probability theory.
Some results of information geometry are shown to remain valid.
Exponential families both of classical and of quantum mechanical statistical physics fit into the new formalism.
Other less obvious applications are predicted.
For instance, quantum states can be modeled as points in a classical phase space
and the resulting model belongs to the exponential family.

\end{abstract}

%%%%%%%%%%%%%%%%%%%%%%%%%%%%%%%%%%%%%%%%%%%%%%%%%%%%%%%%%%%%%%%%%%%%%%%%%%%%%%
\section{Introduction}

The exponential family of statistical models is an important notion in statistics.
The para\-metrized statistical model $\theta\in\Ro^n\rightarrow p_\theta(a)$ belongs to
the {\sl exponential family}\cite {BN78} if there exist functions $\alpha(\theta)$, $c(a)$,
and $H_j(a),j=1,2,\cdots,n$,
such that the probability distributions $p_\theta(a)$ can be written as
\be
p_\theta(a)=c(a)\exp\left[-\alpha(\theta)-\sum_{j=1}^n\theta_jH_j(a)\right].
\label {intro:expfam}
\ee
The choice of signs conforms with the conventions of statistical physics
where the Boltzmann-Gibbs probability distribution is usually written as
\be
p_\beta(a)=\frac {c(a)}{Z(\beta)}e^{-\beta H(a)}.
\ee
This distribution is parametrized by the inverse temperature $\beta$ and
clearly belongs to the exponential family. The function $H(a)$ is called the
Hamiltonian, the normalization $Z(\beta)$ is called the partition sum.
The function $c(a)$ is a prior weight.
In many cases it is identically equal to 1. But for instance, if the underlying measure space $A$ is the set of integers $\No$,
then $c(a)=1/a!$ might be an appropriate choice.

Recently, generalizations of the notion of an exponential family have been
introduced\cite {NJ04,GD04,NJ08,NJ09,OA09,OW09,NJ10,NJ11,AO11}. They provide a solid theoretical
underpinning for research in non-extensive statistical physics\cite{TC88,TC10}.
The equilibrium probability distributions (pdfs) studied in this context are related to
Amari's $\alpha$-family of pdfs\cite {AS85}. The latter is the subject
of research in information geometry\cite {AN00}, where techniques from differential geometry are
applied to probability theory.

The present work has been inspired by the efforts of Tops\o e\cite {TF09,TF11}
to formulate the notion of an exponential family in an abstract setting of game theory.
One of his goals is to formulate information theory without involving statistics.
From \cite{TF09} we quote:
"In 1983 Kolmogorov stated that 'Information theory must precede pobability theory and not be based on it'."
A seminal paper in this direction is the work of Csisz\'ar\cite {CI91}.
The settings of this paper can be reformulated in the terminology used in the present work.
More recent contributions in the area of machine learning are found in \cite {STD08,DV10}.

The next Section introduces the abstract settings of the formalism.
In Section 3 the notion of Entropy is added.
Section 4 gives a definition of an exponential family of models.
Section 5 shows that both the standard and the quantum mechanical notions of an exponential family
fit into the present formalism.
The final Section formulates some conclusions.

%%%%%%%%%%%%%%%%%%%%%%%%%%%%%%%%%%%%%%%%%%%%%%%%%%%%%%%%%%%%%%%%%%%%%%%%%%%%%%
%%%%%%%%%%%%%%%%%%%%%%%%%%%%%%%%%%%%%%%%%%%%%%%%%%%%%%%%%%%%%%%%%%%%%%%%%%%%%%
\section{Data set models}

%%%%%%%%%%%%%%%%%%%%%%%%%%%%%%%%%%%%%%%%%%%%%%%%%%%%%%%%%%%%%%%%%%%%%%%%%%%%%%
\subsection{The information framework}

The elements of our framework are
\begin{description}
 \item The {\sl space of data sets} $\Xo$ is an abstract topological space.
Following Tops\o e \cite {TF09,TF11} an element $x$ of $\Xo$ can be called a {\sl truth}.
However, it is closer to the tradition of probability theory to consider the space of possible
outcomes of an experiment. Therefore we refer to $x$ as a {\sl data set}.
In the probabilistic formulation of information
theory $\Xo$ is the space of pobability distributions over a finite alphabet $A$.
In the quantum mechanical context it is the space of quantum states,
for instance described by normalized wave functions or by density operators.
Other examples are given in what follows.

\item The space of {\sl questions} $\Qo$ is a dual space of $\Xo$. Each question $q$ is a real function
continuously defined on an open subset of $\Xo$.
The evaluation of $q$ in the point $x$ is the {\sl answer to the question} and is denoted
$\langle x|q\rangle$ instead of $q(x)$ to stress that the space of questions is a linear space but not
necessarily an algebra with the usual pointwise product.
For instance, each hermitian bounded operator $A$ on the Hilbert space
of wavefunctions $\psi$ determines an everywhere defined continuous function, given by
\be
\psi\rightarrow \langle\psi|A\rangle\equiv(\psi,A\psi).
\label {dataset:operav}
\ee
Here, $(\phi,\psi)$ is the scalar product of two elements $\phi,\psi$ of the Hilbert space ${\cal L}^2(\Ro^3,\Co)$.
Note that we follow the notational conventions of the physics literature.
In the case of an unbounded operator, such as the position operators or many of the Hamilton
operators, some caution is needed. One must select a topology which makes (\ref  {dataset:operav})
continuous on the domain of definition of the operator.

\end{description}

%\vskip 8pt\noindent

%%%%%%%%%%%%%%%%%%%%%%%%%%%%%%%%%%%%%%%%%%%%%%%%%%%%%%%%%%%%%%%%%%%%%%%%%%%%%%
\subsection{What is a model?}

In statistical physics a model is determined by its Hamiltonian. In the present context
this is replaced by one or more questions. However, we want to make the definition
slightly more general by introducing the following definition.

\begin{definition}
A {\sl data set model} is a topological manifold\footnote{
$\Mo$ is locally Euclidean, this means that there exists in each point $m$ of $\Mo$
an integer $n>0$, an open set $D$ of $\Ro^n$,
together with a map $U\in D\rightarrow x_U\in\Mo$
which is a homeomorphism between $D$ and a neighbourhood of $m$.
}
$\Mo$ together with a continuous map
$\mu$ defined on an open subset of the space $\Xo$ of data sets taking values in $\Mo$.
\end{definition}

Clearly, a set of questions $q_1,\cdots,q_n$ with a common open domain of definition $D$ defines a
manifold $\Mo\subset\Ro^n$ as the range of the map $\mu$ defined by
$\mu(x)=U$ when $U_j=\langle x|q_j\rangle,j=1,2,\cdots,n$, provided that
the set $\mu(D)$ is open in $\Ro^n$.

The converse is also true. Indeed, one has

\begin{proposition}
A local parametrization $U\in D\subset\Ro^n\rightarrow m_U\in\Mo$ of the manifold $\Mo,\mu$
defines questions $q_j$ by $\langle x|q_j\rangle=U_j$ when $\mu(x)=m_U$.
\end{proposition}

\beginproof
The questions are well-defined. The domain of definition is the set
of $x$ for which $\mu(x)$ belongs to the range of the map $U\in D\rightarrow m_U\in\Mo$.
This is an open set because any homeomorphism is an open map. It is also bijective
so that there is a unique $U$ such that $m_U=\mu(x)$. Hence, the answer to the questions $q_j$ is unique.

The map $x\rightarrow\langle x|q_j\rangle=U_j$ is continuous because $\mu(x)$ is continuous and
$U\in D\rightarrow m_U\in\Mo$ is open.

\endproof

The advantage of defining a model in terms of manifolds is that the dependence on a specific choice of questions
has been eliminated.

\begin{example}
The Euclidean space $\Xo=\Ro^3$ is a space of data sets.
The unit sphere
\be
S_2=\{x\in\Ro^3:\,|x|=1\}
\ee
is a model embedded in $\Ro^3$.
The map $\mu$ is defined on $\Ro^3\setminus \{0\}$
by $\mu(x)=x/|x|$.
The questions $q_1$ and $q_2$ defined for $x_3>0$ by
\be
\langle x|q_1\rangle=\frac {x_1}{x_3}
\quad\mbox{ and }\quad
\langle x|q_2\rangle=\frac {x_2}{x_3}.
\ee
determine a parametrization of the northern hemisphere of $S_2$. It is given by
\be
U\rightarrow x_U=(U_1 x_3,U_2 x_3,x_3)^{\rm T}
\quad\mbox{ with }\quad
x_3=\frac 1{\sqrt{1+U_1^2+U_2^2}}.
\ee

\end{example}

%%%%%%%%%%%%%%%%%%%%%%%%%%%%%%%%%%%%%%%%%%%%%%%%%%%%%%%%%%%%%%%%%%%%%%%%%%%%%%
%%%%%%%%%%%%%%%%%%%%%%%%%%%%%%%%%%%%%%%%%%%%%%%%%%%%%%%%%%%%%%%%%%%%%%%%%%%%%%
\section{Maximum entropy principle}

%%%%%%%%%%%%%%%%%%%%%%%%%%%%%%%%%%%%%%%%%%%%%%%%%%%%%%%%%%%%%%%%%%%%%%%%%%%%%%
\subsection{Entropy functions}

The amount of information contained in the data set $x$ is given by its {\sl entropy} $S(x)$.
It is a lower semi-continuous function\footnote{We do not use this property in the present paper.}
with values in the extended reals $[-\infty,+\infty]$.
Usually the entropy is assumed to be concave. However, in general the space $\Xo$ does not have an affine structure.
On the other hand, models are manifolds. Hence, by transferring the notion of entropy to the model points
the concavity as a function of parameters can be discussed.

Given a data set model $\Mo,\mu$ the entropy
$S(m)$ of a model point $m$ is defined by the {\sl maximum entropy principle} of Jaynes\cite {JE57}
\be
S(m)=\sup\{S(x):\,\mu(x)=m\}\le +\infty.
\label{maxent}
\ee
If $m$ is not in the range of $\mu$ then $S(m)=-\infty$ is chosen.
Note that we use here the map $\mu$ as a constraint on the data sets involved in the maximization procedure,
instead of using a specific set of questions $q_1,\cdots,q_n$.

Since $\Mo$ is a manifold we can now investigate whether local parametrizations $U\rightarrow m_U$ exist
such that $S(m_U)$ is a concave function of the parameters $U$.
In what follows the notation $S(U)\equiv S(m_U)$ will be used. Note that $S(U)$ depends on the choice
of local parametrization while $S(m)$ is independent of parametrization.

\begin {proposition}
Let $U\in D\subset\Ro^n\rightarrow m_U$ be a local parametrization of a data set model $\Mo,\mu$,
Let $q_1,\cdots,q_n$ be the accompanying set of questions as defined by Proposition 1.
Then one has locally
\be
S(U)=\sup\{S(x):\,\langle x|q_j\rangle=U_j\mbox{ for }j=1,2,\cdots,n\}\le +\infty.
\ee
\end {proposition}

The proof of this result is straightforward.

\begin{example}
Consider the parametrization of the northern hemisphere of the unit circle, as discussed before.
The entropy function
\be
S(x)=-1-|x|(\ln |x| -1)
\ee
is maximal when $|x|=1$. The entropy function $S(m)$ vanishes on the model manifold.
\end{example}

%%%%%%%%%%%%%%%%%%%%%%%%%%%%%%%%%%%%%%%%%%%%%%%%%%%%%%%%%%%%%%%%%%%%%%%%%%%%%%
\subsection{Perfect data sets}

In the example of the sphere the supremum in (\ref {maxent}) is actually a maximum.
The entropy function $S(x)$ takes on its maximal value for the points of $S_2$. It is then obvious to call
these points {\sl perfect data sets}. Such privileged data points do not always exist. For instance,
the model for a quantum particle can be a point particle localized at a position $q$ in $\Ro^3$.
The map $\mu$ is defined by $\mu(\psi)=\langle\psi|Q\psi\rangle$.
But there are no quadratically integrable wavefunctions which describe a quantum particle perfectly
localized at the position $q$. In such a case one expects an entropy function $S(\psi)$ which is such
that no maximum is attained for any wave function $\psi$.

The relation between model points and perfect data sets may be a one-to-many relation.
This is made clear in the following example.

\begin{example}
In the case of linear regression a data set consists of a finite sequence of pairs of real numbers
\be
(x_1,y_1), (x_2,y_2), \cdots, (x_n,y_n),
\ee
with at least two distinct pairs. The model space consists of straight lines not parallel to the $y$-axis.
A data set is {\sl perfect} if the data points fall on a single line.
But with a single straight line correspond many perfect data sets.
%The map $\xi$ maps the perfect data sets onto the corresponding line.
See the Figure \ref {fig:linregr}.

\begin{figure}[!h!t]
	\centering
	\includegraphics[height=4cm]{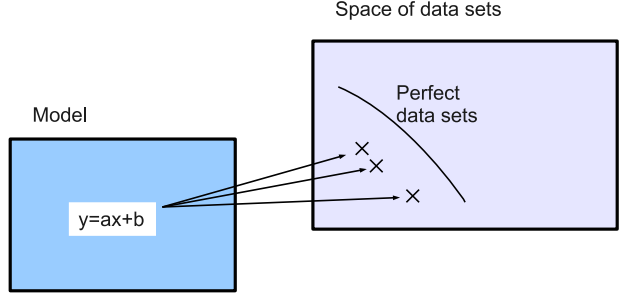}
	\caption{Embedding of the model into the space of data sets.}
	\label {fig:linregr}
\end{figure}

% 
% In the case of a perfect data set there exist real numbers $a$ and $b$ such that
% \be
% y_i=ax_i+b,\qquad i=1,\cdots,n.
% \ee
The interesting questions are given by
\begin{itemize}
 \item $\displaystyle q_a(x,y)=\frac 1{Z}\sum_{i,j}(y_i-y_j)(x_i-x_j)$;
 \item $\displaystyle q_b(x,y)=\frac 1{Z}\sum_{i,j}\sum_{i,j}(x_iy_j-x_jy_i)(x_i-x_j)$.
\end{itemize}
with $Z=\sum_{i,j}(x_i-x_j)^2$.
They are only defined on data sets for which $Z\not=0$.
They are interesting because they return the parameters $a$ and $b$ of the fitted line $y=ax+b$.
These two questions uniquely determine the model.
A meaningful entropy function is
\be
S(x,y)=-\frac 1{Z}\sum_{i,j=1}^n(x_iy_j-x_jy_i)^2-\frac 1{Z}\sum_{i,j}^n(y_i-y_j)^2.
\ee
Its value on perfect data sets is $-a^2-b^2$. For other data sets is $S(x)<S(\mu(x))$.

\end{example}

%%%%%%%%%%%%%%%%%%%%%%%%%%%%%%%%%%%%%%%%%%%%%%%%%%%%%%%%%%%%%%%%%%%%%%%%%%%%%%
%%%%%%%%%%%%%%%%%%%%%%%%%%%%%%%%%%%%%%%%%%%%%%%%%%%%%%%%%%%%%%%%%%%%%%%%%%%%%%
\section{Exponential families}

The notion of an exponential family of models is strongly related to the concept
of canonical parametrizations. These are introduced now.

%%%%%%%%%%%%%%%%%%%%%%%%%%%%%%%%%%%%%%%%%%%%%%%%%%%%%%%%%%%%%%%%%%%%%%%%%%%%%%
\subsection{Contact transforms}

In thermodynamics, the {\sl Massieu function} $\Phi(\theta)$
is the Legendre transform of the entropy $S(U)$.
This inspires for the following definition.

\begin{definition}
Let be given a local parametrization 
$U\in D\subset\Ro^n\rightarrow m_U$
of a data set model $\Mo,\mu$.
Assume that the model entropy $S(U)$ is locally finite.
Then the Massieu function is defined by
\be
\Phi(\theta)=\sup_{U\in D}\left\{S(U)-\sum_{j=1}^n\theta_jU_j\right\}.
\label {massieu}
\ee
\end{definition}

\begin{theorem}
Let be given a local parametrization 
$U\in D\subset\Ro^n\rightarrow m_U$
of a data set model $\Mo,\mu$. Let $q_1,\cdots,q_n$ be the accompanying set of questions 
defined by Proposition 1.
Assume that the model entropy $S(U)$ is locally finite.
Then one has
 \be
\Phi(\theta)=\sup\{S(x)-\sum_{j=1}^n\theta_j\langle x|q_j\rangle:\,\mu(x)\mbox{ is local}\}.
\label {varprin}
\ee
 $\Phi(\theta)$ is a convex function. In particular, it is finite on a convex subset $\Theta$
of $\Ro^n$.
% On the interior of $\Theta$ is $\Phi(\theta)$ strictly convex and differentiable with
% \be
% \frac{\partial\Phi}{\partial\theta_j}=-U_j,
% \ee
% with $U$ related to $\theta$ by the requirement that $\Phi(\theta)=S_q(U)-\sum_{j=1}^n\theta_jU_j$.
\end{theorem}

\beginproof
Remember that the questions are such that $\mu(x)=m_U$ holds if and only if $\langle x|q_j\rangle=U_j$ for $j=1,2,\cdots,n$.
Take $x$ so that $\mu(x)$ is local. Then one has $\mu(x)=m_U$ with $U\in D$.
But $S(U)<+\infty$ implies that $S(x)<+\infty$. Hence one has
\be
S(x)-\sum_{j=1}^n\theta_j\langle x|q_j\rangle\le S(U)-\sum_{j=1}^n\theta_jU_j\le \Phi(\theta).
\ee

On the other hand, if $\Phi(\theta)<+\infty$ then for any $\epsilon>0$ there exists $U$ such that
\be
\Phi(\theta)-\epsilon<S(U)-\sum_{j=1}^n\theta_jU_j.
\ee
Similarly, there exists $x$, satisfying $\langle x|q_j\rangle=U_j$ for $1=1,2,\cdots,n$,
such that
\be
S(U)-\epsilon<S(x).
\ee
All together one has
\be
\Phi(\theta)-2\epsilon<S(x)-\sum_{j=1}^n\theta_jU_j.
\ee
Since $\epsilon>0$ is arbitrary one concludes that the equality holds in (\ref {varprin}).

Finally, if $\Phi(\theta)=+\infty$ then there exists $U$ such that $S(U)-\sum_{j=1}^n\theta_jU_j$ is arbitrary large.
But then there exists $x$ such that $\mu(x)$ is local and $S(x)-\sum_{j=1}^n\theta_j\langle x|q_j\rangle$ is arbitrary large.
Hence, also in this case the equality holds in (\ref {varprin}).

The convexity statement is easy to show. Let $\lambda$ in $[0,1]$.
One can assume that $\Phi(\theta_1)$ and $\Phi(\theta_2)$ are finite because otherwise the convexity statement is empty.
Then for any $x$ with local $\mu(x)$ one has
\be
& &S(x)-\sum_{j=1}^n[\lambda\theta_{1,j}+(1-\lambda)\theta_{2,j}]\langle x|q_j\rangle\cr
&=&\lambda\left[S(x)-\sum_{j=1}^n\theta_{1,j}\langle x|q_j\rangle\right]
+(1-\lambda)\left[S(x)-\sum_{j=1}^n\theta_{2,j}\langle x|q_j\rangle\right]\cr
&\le&
\lambda\Phi(\theta_1)+(1-\lambda)\Phi(\theta_2).
\ee
This implies $\Phi(\lambda\theta_1+(1-\lambda)\theta_2)\le \lambda\Phi(\theta_1)+(1-\lambda)\Phi(\theta_2)$.

\endproof

In the physics literature one is used to work with the free energy
rather than with Massieu's function. If the inverse temperature $\beta$ is the only parameter
then the free energy equals $-\Phi(\beta)/\beta$
and minimizes $\langle x|q\rangle- S(x)/\beta$.

%%%%%%%%%%%%%%%%%%%%%%%%%%%%%%%%%%%%%%%%%%%%%%%%%%%%%%%%%%%%%%%%%%%%%%%%%%%%%%
\subsection{Canonical parametrization}

Let us now return to a data set model with a locally defined parametrization.
Then the Legendre-Fenchel transform can be used to introduce a canonical parametrization.
The attribute 'canonical' refers to the canonical ensemble of statistical physics.
In the context of the exponential family one speaks about the canonical form of the probability distribution.
But in the present approach the canonical parametrization is defined before introducing the exponential family
and is independent of it.

\begin{definition}
Let be given some local parametrization $\theta\in\Theta\subset\Ro^n\rightarrow m_\theta$
of a data set model $\Mo,\mu$. 
%Assume $\Theta$ is an open set of $\Ro^n$.
The parametrization is said to be {\sl canonical}
if there exists another local parametrization
$U\in D\subset\Ro^n\rightarrow m_U$ such that
\begin{itemize}
 \item $S(U)<+\infty$ for all $U$ in $D$;
 \item The relation $m_\theta=m_U$ defines a diffeomorphism between $\Theta$ and $D$;
 \item Under this diffeomorphism is
\be
\Phi(\theta)-S(U)+\sum_{j=1}^n\theta_jU_j=0.
\label {canon:identity}
\ee
\end{itemize}

\end{definition}

To make the distinction between the two parametrizations $\theta\in\Theta\subset\Ro^n\rightarrow m_\theta$
and $U\in D\subset\Ro^n\rightarrow m_U$ we call the latter the associated energy parametrization.
The motivation is that in statistical physics the components of $U$ have the meaning of energies.

\begin{theorem}
If the parametrization $\theta\in\Theta\rightarrow m_\theta$
of a data set model $\Mo,\mu$ is canonical then the Massieu function $\Phi(\theta)$ is a strictly convex
differentiable function and there exist questions  $q_1,\cdots,q_n$ satisfying
\be
\frac {\partial\,}{\partial\theta_j}\Phi(\theta)
=-\langle x|q_j\rangle
\quad\mbox{ for all } x \mbox{ satisfying }\mu(x)=m_\theta.
\label {phideriv}
\ee
\end{theorem}

\beginproof
Let $U\in D\subset\Ro^n\rightarrow m_U$ be the local parametrization appearing
 in the definition of a canonical parametrization.
Note that
\be
\zeta\rightarrow \Phi(\theta)-\sum_{j=1}^nU_j(\zeta_j-\theta_j)
\ee
is a tangent plane in the point $\theta$.
The requirement that $m_U=m_\theta$ determines a diffeomorphism implies that a small
change of $\theta$ corresponds with a small change of $U$
and hence a small change in the slope of the tangent plane. This proves that the tangent plane is unique.
One concludes that $\Phi(\theta)$ is differentiable and that
\be
\frac {\partial\Phi}{\partial\theta_j}
=-U_j.
\ee
The strict convexity follows because the correspondence $\theta\leftrightarrow U$ is bijective.

Let $q_1,\cdots,q_n$ be the questions defined in Proposition 1.
They satisfy $\langle x|q_j\rangle=U_j$ for $j=1,2,\cdots,n$ when $\mu(x)=m_U$.
Hence the statement of the Theorem follows.

\endproof

The second derivatives of $\Phi(\theta)$ define a metric tensor
\be
g_{j,k}(\theta)=\frac {\partial^2\Phi}{\partial\theta_j\partial\theta_k}=-\frac{\partial U_k}{\partial \theta_j}.
\label {canonical:metrictensor}
\ee
This matrix is a generalization of Fisher's information matrix.

\begin{example}
Let $\Xo$ be the set of all 2-by-2 density operators
(these are positive trace class operators with trace equal to 1).
The entropy function is the von Neumann entropy
\be
S(\rho)=-\Tr\rho\ln\rho.
\label {vNentropy}
\ee
The model $\Mo$ coincides with the space of data sets $\Xo$.
Let us calculate a  parametrization which is canonical.

Three questions are needed to determine uniquely a density operator $\rho$.
In terms of the three Pauli matrices $\sigma_j$ these are
\be
\langle\rho|q_j\rangle=\Tr\rho\sigma_j, \quad j=1,2,3.
\label {quexquest}
\ee 
Then one can write
\be
\rho=\frac 12\left(\Io+\sum_jU_j\sigma_j\right)
\quad\mbox{ with }
U_j=\langle\rho|q_j\rangle.
\ee
The von Neumann entropy becomes
\be
S(\rho)=\ln 2-\frac 12(1+|U|)\ln (1+|U|)-\frac 12(1-|U|)\ln (1-|U|).
\ee
The Massieu function reads
\be
\Phi(\theta)&=&\sup_{U}\{S(U)-\sum_{j=1}^3\theta_jU_j:\, |U|\le 1\}.\cr
& &
\ee
The maximum is reached when
 \be
 \theta_j=\frac 12\frac {U_j}{|U|}\ln\frac {1-|U|}{1+|U|}.
\ee
Note that this implies that $|U|=\tanh|\theta|$.
Hence the inverse relation is
\be
U_j=-\frac {\theta_j}{|\theta|}\tanh|\theta|.
\ee
One concludes that the map $U\rightarrow\theta$ is a diffeomorphism from the interior of the unit sphere
onto $\Ro^3$.

$\rho_\theta$ can now be written as
\be
\rho_\theta
&=&\frac 12\Io-\frac 1{2|\theta|}\tanh|\theta|\sum_{j=1}^3\theta_j\sigma_j\cr
&=&\frac 1{2\cosh(|\theta|)}\exp\left(-\sum_j\theta_j\sigma_j\right).
\label {quexample}
\ee
This is a canonical parametrization of the 2-by-2 density matrices.

\end{example}

%%%%%%%%%%%%%%%%%%%%%%%%%%%%%%%%%%%%%%%%%%%%%%%%%%%%%%%%%%%%%%%%%%%%%%%%%%%%%%
\subsection{Dual Relations}

Let be given a canonical parametrization $\theta\rightarrow m_\theta$ of
model $\Mo,\mu$, together with the associated energy parametrization
$U\rightarrow m_U$. From (\ref {canon:identity}, \ref {phideriv}) then follows
the pair of dual relations
\be
\frac {\partial\Phi}{\partial\theta_j}=-U_j
\quad\mbox{ and }\quad
\frac {\partial S}{\partial U_j}=\theta_j,
\ee
where $U\rightarrow \theta$ is the diffeomorphism determined by the relation $m_U=m_\theta$.

The function $S(U)$ is strictly concave. This follows because
the matrix of second derivatives of $S(U)$ equals minus the
inverse of the metric tensor $g_{j,k}(\theta)$ defined by (\ref {canonical:metrictensor}).
The latter is positive definite because by Theorem 2 the Massieu function is strictly convex.

If the metric tensor $g_{j,k}(\theta)$ is sufficiently smooth then the model space $\Mo$
is (locally) a Riemannian manifold with respect to each of the two parametrizations.
They are dual to each other in the sense that the metric tensor of one parametrization
is the inverse of that of the other. The curvature of the manifold in the Levi-Civita connection
 vanishes because the
metric tensor is the matrix of second derivatives of a convex function.
Hence the manifold is flat.

%%%%%%%%%%%%%%%%%%%%%%%%%%%%%%%%%%%%%%%%%%%%%%%%%%%%%%%%%%%%%%%%%%%%%%%%%%%%%%
\subsection{Logarithmic maps}

%Fix a data set model $\Mo,\mu$.

\begin{definition}
A {\sl logarithmic map} $L$ maps model points onto questions.
% in such a way that
% for all model points $m$ the value of $\langle x|Lm\rangle$ is constant on the set of $x$ for which $\mu(x)=m$.

% \be
% \langle x|Lx-Ly\rangle\ge \langle y|Lx-Ly\rangle
% \quad\mbox{ for all }x,y \in\Xo.
% \ee
\end{definition}

For instance, the Boltzmann-Gibbs-Shannon entropy $S(p)$ can be written as the average of the
measurable quantity $-\ln p(i)$. The probability distribution $p$ belongs to the
space of data sets $\Xo$. But  $-\ln p(i)$ is used as a question, the answer of which
is the value of the entropy function $S(p)$. In this example the logarithmic map is defined on all
data sets. But we need it further on only for perfect data sets or for model points.

 The logarithmic map $L$ can be used to define a {\sl divergence} or {\sl relative entropy}
between data sets and model points.

\begin{definition}
The divergence of a data set $x$ from a model point $m$ is given by
\be
D(x||m)=\sup\{S(y)+\langle y|Lm\rangle:\,\mu(y)=m\}-S(x)-\langle x|Lm\rangle.
\ee
\end{definition}

Clearly, if $\mu(x)=m$ then $D(x||m)\ge 0$ with equality if and only if $x$ maximizes
$S(x)+\langle x|Lm\rangle$ under the constraint $\mu(x)=m$. We call such $x$
{\sl canonical} data sets.

%%%%%%%%%%%%%%%%%%%%%%%%%%%%%%%%%%%%%%%%%%%%%%%%%%%%%%%%%%%%%%%%%%%%%%%%%%%%%
\subsection{Exponential families}

In the previous subsection the notion of a logarithmic map was introduced to prepare for the
definition of the exponential family.

\begin{definition}
A model $\Mo,\mu$ with logarithmic map $L$ belongs to the {\sl exponential family}
of data set models if the model space $\Mo$ is covered with local parametrizations
$\theta\in\Theta\rightarrow m_\theta$,
which are canonical,
and the associated energy parametrizations $U\in D\subset\Ro^n\rightarrow m_U$  are such that
\be
Lm_\theta=\alpha(\theta)-\sum_j\theta_jq_j
\quad\mbox{ for all }\theta\in\Theta,
\label {exp:exp}
\ee
where the questions $q_j$ are defined by $\langle x|q_j\rangle=U_j$ when $\mu(x)=m_U$ (see Proposition 1).

\end{definition}

In the example of the 2-by-2 density matrices (see (\ref {quexample})) is
\be
\ln \rho_\theta=-\ln {2\cosh(|\theta|)}-\sum_j\theta_j\sigma_j.
\ee
Hence the model belongs to the exponential family. One has $\alpha(\theta)=-\ln {2\cosh(|\theta|)}$.
The questions $q_j$ are given by (\ref {quexquest}).

The property (\ref {exp:exp}) can be used to simplify the Definition 5 of divergence.
One obtains
\be
D(x||m_\theta)
&=&\sup\{S(y)+\langle y|\alpha(\theta)-\sum_j\theta_jq_j\rangle:\,\mu(y)=m_\theta\}\cr
& &-S(x)-\langle x|\alpha(\theta)-\sum_j\theta_jq_j\rangle\cr
&=&\sup\{S(y)-\langle y|\sum_j\theta_jq_j\rangle:\,\mu(y)=m_\theta\}\cr
& &-S(x)+\sum_j\theta_j\langle x|q_j\rangle\cr
&=&\Phi(\theta)-S(x)+\sum_j\theta_j\langle x|q_j\rangle.
\label {exp:simpl}
\ee
From Theorem 1 now follows that $D(x||m_\theta)\ge 0$ for all $x$ for which $\mu(x)$ is local.
Equality then holds if and only if the data set is canonical.

Note that one can write, using (\ref {canon:identity}),
\be
D(x||m_\theta)
&=&S(U)-\sum_j\theta_jU_j-\left[S(x)-\sum_j\theta_j\langle x|q_j\rangle\right].
\ee
If $\mu(x)=m_U$ then $\langle x|q_j\rangle=U_j$. Hence
\be
D(x||m_\theta)&=&S(U)-S(x)\ge 0
\quad\mbox{ if }\mu(x)=m_U.
\ee
Therefore, in the case of a model belonging to the exponential family,
 canonical data sets are perfect data sets as well.

%%%%%%%%%%%%%%%%%%%%%%%%%%%%%%%%%%%%%%%%%%%%%%%%%%%%%%%%%%%%%%%%%%%%%%%%%%%%%
\subsection{Pythagorean Theorems}

The model map $\mu$ can be seen as an orthogonal projection of $\Xo$ onto the manifold $\Mo$.
This is supported by a Pythagorean theorem in which the divergence plays the role of a
distance squared.

Introduce the divergence between two model points $m$ and $m'$ by
\be
D(m||m')=\inf\{D(x||m'):\,\mu(x)=m\}.
\ee
The following result shows that this divergence is of the Bregman type\cite {CI91,BLM67}. It has a nice
geometric interpretation. It is the difference between the value $\Phi(\zeta)$
of the Massieu function in the point $\zeta$ and the value of the plane tangent
in the point $\theta$.

\begin{proposition}
Let be given a model $\Mo,\mu$ with logarithmic map $L$ belonging to the exponential family.
Consider a local parametrization $\theta\in\Theta\rightarrow m_\theta$
and the associated energy parametrization $U\in D\subset\Ro^n\rightarrow m_U$
as in the definition of the exponential family.
Then one has
\be
D(m_\theta||m_\zeta)
&=&\Phi(\zeta)-\Phi(\theta)+\sum_j(\zeta_j-\theta_j)U_j.
\label {pyth:modeldist}
\ee
\end{proposition}

\beginproof
First calculate using (\ref {exp:simpl})
\be
D(m_\theta||m_\zeta)
&=&\inf\{D(y||m_\zeta):\,\mu(y)=m_\theta\}\cr
&=&\Phi(\zeta)-\sup\{S(y)-\sum_j\zeta_j\langle y|q_j\rangle:\,\mu(y)=m_\theta\}.
\ee
Now use that $\langle y|q_j\rangle$ is constant on the set of $y$ for which $\mu(y)=m_\theta$.
Hence one has
\be
D(m_\theta||m_\zeta)
&=&\Phi(\zeta)-S(U)+\sum_j\zeta_jU_j
\ee
with $U$ so that $m_U=m_\theta$.
Using (\ref {canon:identity}) this becomes (\ref {pyth:modeldist}).

\endproof

\begin{figure}[!h!t]
	\centering
	\includegraphics[height=6cm]{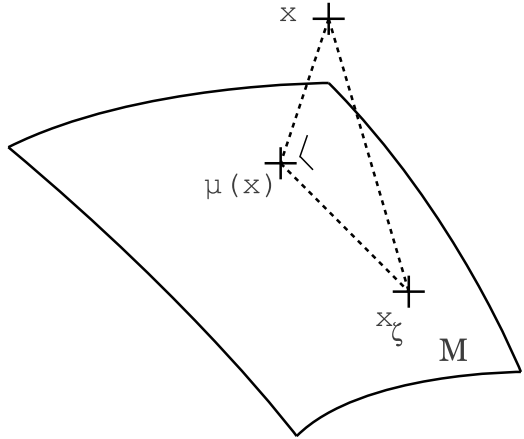}
	\caption{Projection of a data set $x$ onto the manifold $\Mo$ using the model map $\mu$.}
	\label {fig:phyt}
\end{figure}

The Pythagorean theorem\cite {CI91} for the projection of an arbitrary data set $x\in\Xo$ onto the manifold $\Mo$
by means of the model map $\mu$ now follows readily. See the Figure \ref {fig:phyt}.

\begin{theorem}
Let be given a model $\Mo,\mu$ with logarithmic map $L$ belonging to the exponential family.
If $\mu(x)=m_\theta$ then
\be
D(x||m_\theta)+D(m_\theta||m_\zeta)
&=&D(x||m_\zeta).
\label {pyth:thm1}
\ee
\end{theorem}

\beginproof
Use (\ref {pyth:modeldist}) to obtain
\be
D(x||m_\theta)+D(m_\theta||m_\zeta)
=\Phi(\zeta)-S(x)+\sum_j\zeta_j\langle x|q_j\rangle
=D(x||m_\zeta).
\ee
This is (\ref {pyth:thm1}).

\endproof

Following \cite {AO11}, we can also formulate a Pythagorean theorem involving only
model points. 

\begin{theorem}
Consider a model $\Mo,\mu$ with logarithmic map $L$ belonging to the exponential family.
Let $\theta\in\Theta\rightarrow m_\theta$ and $U\in D\rightarrow m_U$ be canonical and energy parametrizations
as mentioned in Definition 6.
Let $\theta,\zeta,\xi$ be points in $\Theta$. Let $U$ and $V$ be dual coordinates such that $m_U=m_\theta$ and $m_V=m_\zeta$.
Assume that
\be
\sum_j(\zeta_j-\xi_j)(U_j-V_j)=0.
\ee
Then one has
\be
D(m_\theta||m_\zeta)+D(m_\zeta||m_\xi)=D(m_\theta||m_\xi).
\ee
\end{theorem}

\beginproof
This follows immediately from (\ref {pyth:modeldist}).
\endproof

%%%%%%%%%%%%%%%%%%%%%%%%%%%%%%%%%%%%%%%%%%%%%%%%%%%%%%%%%%%%%%%%%%%%%%%%%%%%%%
%%%%%%%%%%%%%%%%%%%%%%%%%%%%%%%%%%%%%%%%%%%%%%%%%%%%%%%%%%%%%%%%%%%%%%%%%%%%%%
\section{Applications}

We show below how the standard notion of an exponential family of statistical models
fits into the present formalism. Also the analogue notion in quantum statistics is discussed.
The generalized exponential families\cite {NJ04} 
introduced in the context of Tsallis' non-extensive statistical mechanics\cite{TC10},
or even in a broader context, do fit as well, but will not be treated here.

%%%%%%%%%%%%%%%%%%%%%%%%%%%%%%%%%%%%%%%%%%%%%%%%%%%%%%%%%%%%%%%%%%%%%%%%%%%%%%
\subsection{Statistical models}

Here we show that the above framework is a generalization of
the notion of the exponential family of statistical models\cite {BN78}.

Let $\Xo$ be the affine space of probability distributions over the discrete measure space $A$. Let
$c(a)$ be a prior weight on $A$.
Questions are real functions $f$ of $A$,
seen as maps $p\rightarrow\Eo_p f=\sum_ap(a)f(a)$. The answer to a question $f$, given $p$, is therefore given by
\be
\langle p|f\rangle=\Eo_pf.
\ee
The entropy function is that of Boltzmann-Gibbs-Shannon (BGS) and is given by
\be
S(p)=-\langle p|L(p)\rangle=-\sum_ap(a) \ln  \frac {p(a)}{c(a)}.
\ee

Let $\theta\in\Theta\rightarrow p_\theta$ be a statistical model with probability distributions $p_\theta$
given by (\ref {intro:expfam}). 
For convenience assume $c(a)=1$ and introduce the notation $\Eo_\theta\equiv\Eo_{p_\theta}$.
Let $U_j(\theta)=\Eo_\theta H_j$. The model space $\Mo$ is the subset of $\Xo$ given by
\be
\Mo=\{p_\theta:\theta\in\Theta\}.
\ee
Introduce the model map $\mu$ by
\be
\mu(p)=p_\theta\quad\mbox{ if }\Eo_pH_j=\Eo_\theta H_j\mbox{ for }j=1,\cdots,n.
\ee
Assume for convenience that the functions $H_j(a)$ are bounded. Then the model map is everywhere
defined and continuous in the $l_1$-metric of $\Xo$.

It is well-known that the probability distributions of a model belonging to the exponential family
maximise the BGS-entropy under the constraint $U_j(\theta)=\Eo_\theta H_j$, $j=1,\cdots,n$
--- in our terminology the $p_\theta$ are perfect data sets.
Hence one has
\be
S(\theta)=S(U)=S(p_\theta)=\alpha(\theta)+\sum_j\theta_jU_j(\theta).
\label {appl:statentr}
\ee
In particular, there follows that $\Phi(\theta)=\alpha(\theta)$.

Generically, the relation between $U$ and $\theta$ is a diffeomorphism. Indeed, one has
\be
g_{j,k}(\theta)&=&-\frac {\partial U_k}{\partial \theta_j}\cr
&=&-\frac {\partial\,}{\partial \theta_j}\sum_ap_\theta(a)H_k(a)\cr
&=&\sum_ap_\theta(a)H_j(a)H_k(a)+\sum_ap_\theta(a)\frac {\partial\alpha}{\partial\theta_j}H_j\cr
&=&\Eo_\theta H_jH_k-\left(\Eo_\theta H_j\right)\left(\Eo_\theta H_k\right).
\ee
If the constant function is not a linear combination of the hamiltonians $H_j$ then the matrix $g_{j,k}(\theta)$ is positive definite.
This implies that the relation between $U$ and $\theta$ is a diffeomorphism.

One concludes that the parametrization $\theta\rightarrow p_\theta$ is canonical.

Introduce a logarithmic map $L$ by
\be
(Lp_\theta)(a)=\ln p_\theta(a).
\ee
The corresponding divergence is
\be
D(p||p_\theta)
&=&\sum_ap(a) \ln  \frac {p_\theta(a)}{p(a)}.
\ee
This is the standard expression for the divergence/relative entropy.

It follows now from (\ref {appl:statentr}) that the model $\Mo,\mu$ with this logarithmic map
belongs to the exponential family provided that no linear combination of the hamiltonians $H_j$ is a constant function.

%%%%%%%%%%%%%%%%%%%%%%%%%%%%%%%%%%%%%%%%%%%%%%%%%%%%%%%%%%%%%%%%%%%%%%%%%%%%%%
\subsection{Quantum statistical physics}

In quantum statistics the probability distributions of classical statistics are replaced
by density matrices/density operators on a separable Hilbert space.
They form the space $\Xo$ of data sets. Questions are bounded operators on the Hilbert space.
The evaluation function is
\be
\rho\in\Xo\rightarrow \langle \rho|A\rangle\equiv\Tr\rho A.
\ee
It is continuous for instance in the Hilbert-Schmidt norm.
The entropy function is the von Neumann entropy (\ref {vNentropy}).

A quantum statistical model is a homeomorphism $\theta\in\Theta\subset\Ro^n\rightarrow\rho_\theta$.
The model space is $\Mo=\{\rho_\theta:\,\theta\in\Theta\}$.
The model bolongs to the exponential family of quantum models if there exist
self-adjoint operators $H_1,\cdots,H_n$ such that
\be
\rho_\theta=\frac 1{Z(\theta)}\exp(-\sum_{j=1}^n\theta_jH_j)
\label {qsp:expfam}
\ee
with $Z(\theta)=\Tr \exp(-\sum_{j=1}^n\theta_jH_j)$.
The model map $\mu$ satisfies $\mu(\rho)=\rho_\theta$ if $\Tr\rho H_j$ is well-defined
and equals $U_j=\Tr\rho_\theta H_j$ for $j=1,\cdots,n$.

The $\rho_\theta$ of the form (\ref {qsp:expfam}) maximize the von Neumann entropy 
under the constraint of a given value of the $U_j$.
The proof is based on Klein's inequality --- see for instance \cite {RD69,NJ11}.
In particular the $\rho_\theta$ are perfect data sets. One obtains
\be
S(U)=S(\rho_\theta)=\Phi(\theta)+\sum_{j=1}^n\theta_jU_j
\quad\mbox{ with }\quad\Phi(\theta)=\ln Z(\theta).
\ee

One calculates
\be
g_{j,k}(\theta)&=&-\frac {\partial U_k}{\partial \theta_j}
=-\frac {\partial \,}{\partial \theta_j}\Tr\rho H_k\cr
&=&\Tr\rho H_jH_k-\frac {\partial Z}{\partial \theta_j}\Tr\rho H_k\cr
&=&\Tr\rho H_jH_k-(\Tr\rho H_j)(\Tr\rho H_k).
\ee
The eigenvalues of this matrix cannot be negative.
If they are strictly positive for all $\theta$ then the relation between $U$ and $\theta$
is a diffeomorphism and the parametrization $\theta\rightarrow \rho_\theta$ is canonical.

Introduce the logarithmic map defined by $L\rho_\theta=\ln \rho_\theta$.
One clearly has
\be
L\rho_\theta=-\ln Z(\theta)-\sum_{j=1}^n\theta_jH_j.
\ee
Hence, the model belongs to the exponential family according to Definition 6.
A short calculation then yields
\be
D(\rho||\rho_\theta)&=&\Tr\rho(\ln\rho-\ln\rho_\theta).
\ee
This is the standard expression for relative entropy in quantum statistical physics\cite {PD07}.

%%%%%%%%%%%%%%%%%%%%%%%%%%%%%%%%%%%%%%%%%%%%%%%%%%%%%%%%%%%%%%%%%%%%%%%%%%%%%%
\subsection{Coherent states}

Now we discuss an example which shows that our framework extends well beyond the (quantum) statistical context.
We consider the phase space of classical mechanics as a model for a state space of quantum mechanical
wave functions.

For simplicity consider a quantum particle in one dimension. The space $\Xo$ of data sets consists of
wave functions $\psi(x)$ which are twice differentiable and normalized so that
\be
\int_\Ro\upd x\,|\psi(x)|^2=1.
\ee
Note that two wave functions $\psi(x)$ and $e^{i\alpha\psi(x)}$, with $\alpha$ constant, determine the same point of $\Xo$.

Questions are linear operators $A$ acting on the Hilbert space of square integrable complex functions.
The evaluation function is given by
\be
\langle \psi|A\rangle=\int_\Ro\upd x\,\overline{\psi(x)}(A\psi)(x).
\ee
Introduce position and momentum operators by $Q\psi(x)=x\psi(x)$ and
$\displaystyle P\psi(x)=-i\hbar\frac {\partial\psi}{\partial x}$.
Note that these are unbounded operators. Hence we need a topology on $\Xo$
which is such that the two questions $\psi\rightarrow \langle \psi|Q\rangle$
and $\psi\rightarrow \langle \psi|P\rangle$ are continuous.
Then they define a continuous map $\mu$ of $\Xo$ into the model space $\Mo=\Ro^2$, which is the phase space
of a particle in classical mechanics.

Introduce now the entropy function
\be
S(\psi)=\frac 12\left|\langle\psi|a\rangle\right|^2-\langle\psi|a^\dagger a\rangle,
\label{entropy_coherent}
\ee
where the annihilation operator $a$ is defined by
\be
a=\frac 1{\sqrt 2}\left[\frac 1r Q+i\frac r\hbar P\right],
\ee
with $r$ and $\hbar$ positive constants.
Then $\Xo$ together with this entropy function is a data set space.

The solution of the eigen equation $a\psi=z\psi$, with complex $z$, is denoted $\psi_z$ and is called a
{\sl coherent state}. 
Note that
\be
U_1=\langle \psi_z|Q\rangle=r\sqrt 2\,\Re z
\quad\mbox{ and }\quad
U_2=\langle \psi_z|P\rangle=\frac {\hbar}r \sqrt 2\,\Im z.
\label {coherent:Uz}
\ee
and
\be
|\langle\psi|a\rangle|^2=\frac 1{2r^2}U_1^2+\frac {r^2}{2\hbar^2}U_2^2.
\ee
Clearly is
\be
S(\psi_z)=-\frac 12\left|\langle\psi_z|a\rangle\right|^2=-\frac 12|z|^2,
\ee
and
\be
S(\psi)\le -\frac 12\left|\langle\psi|a\rangle\right|^2
\quad\mbox{ for all }\psi\in\Xo
\mbox{ for which }
\langle\psi|a\rangle=z.
\ee
Hence, the coherent states are perfect data sets. In particular, the entropy $S(m)$ of the model point $m=m_U$ is
\be
S(U)=-\frac 1{2r^2}U_1^2-\frac {r^2}{2\hbar^2}U_2^2.
\ee
The Massieu function equals
\be
\Phi(\theta)=\sup_U\{S(U)-\theta_1U_1-\theta_2U_2\}.
\ee
The maximum is reached when
\be
\theta_1=-\frac 1{r^2}U_1
\quad\mbox{ and }\quad\theta_2=-\frac {r^2}{\hbar^2}U_2.
\ee
The result is
\be
\Phi(\theta)=\frac {r^2}{2}\theta_1^2+\frac {\hbar^2}{2r^2}\theta_2^2.
\ee
It is now straightforward to verify that the $\theta$-parametrization of $\Ro^2$ is canonical.

Introduce a logarithmic map $L$ by
\be
L(m_U)
&=&-\frac 12|z|^2+\frac 12z a^\dagger+\frac 12\overline z a,
\ee
where $z$ is obtained from (\ref {coherent:Uz}).
There follows immediately that
\be
L(m_U)
&=&-\Phi(\theta)-\theta_1 Q- \theta_2 P.
\ee
This shows that the model belongs to the exponential family.
The divergence equals
\be
D(\phi||m_U)
&=&\frac 12|\langle\phi|a\rangle-z|^2+\langle\phi| a^\dagger a\rangle-|\langle\phi| a\rangle|^2
\ge 0.
\ee
In addition, $D(\phi||\psi_z)=0$ is equivalent with $z=\langle\phi|a\rangle$ and
$a\phi=\langle\psi| a\rangle\phi$. But this implies that $\phi$ equals $\psi_z$,
up to a phase factor which can be neglected because it has no physical meaning.
Hence, the divergence vanishes if and only if $\phi$ equals $\psi_z$ up to a constant phase factor.

%%%%%%%%%%%%%%%%%%%%%%%%%%%%%%%%%%%%%%%%%%%%%%%%%%%%%%%%%%%%%%%%%%%%%%%%%%%%%%
%%%%%%%%%%%%%%%%%%%%%%%%%%%%%%%%%%%%%%%%%%%%%%%%%%%%%%%%%%%%%%%%%%%%%%%%%%%%%%
\section{Conclusions}

The notion of an exponential family of models can be generalized to a context
not involving probability theory. From the point of view of statistical physics
this is of interest because the exponential family is at the heart of the discipline
and quantum statistical physics involves quantum probability rather than classical
probability theory. But the formalism presented here is so general that it has
many other applications. Only one such example has been elaborated in subsection 5.3.
Some other applications have been mentioned without proof. These will be taken up
in further work.

By the present effort we hope to contribute to a more general theory of information, including
previous extensions in the directions of machine learning, statistical inference and
quantum information.
%  Our motivation comes from physics. The exponential family of quantum
% models can be formulated in analogy to the standard notion of an exponential family.
% But it is clearly not situated within classical probability theory. The formalism presented
% in the present paper succeeds to bring the classical and quantum formulations together in a
% single theory.

%%%%%%%%%%%%%%%%%%%%%%%%%%%%%%%%%%%%%%%%%%%%%%%%%%%%%%%%%%%%%%%%%%%%%%%
%%%%%%%%%%%%%%%%%%%%%%%%%%%%%%%%%%%%%%%%%%%%%%%%%%%%%%%%%%%%%%%%%%%%%%%
%%%%%%%%%%%%%%%%%%%%%%%%%%%%%%%%%%%%%%%%%%%%%%%%%%%%%%%%%%%%%%%%%%%%%%%
%%%%%%%%%%%%%%%%%%%%%%%%%%%%%%%%%%%%%%%%%%%%%%%%%%%%%%%%%%%%%%%%%%%%%%%


\begin{thebibliography}{99}
\parskip 0pt\raggedright\small

\bibitem {BN78}
O. E. Barndorff-Nielsen, {\sl Information and Exponential Families in Statistical Theory}
(J. Wiley and Sons, New York, 1978).

\bibitem {NJ04}
 J. Naudts, {\sl Estimators, escort probabilities, and phi-exponential families in statistical physics,}
J. Ineq. Pure Appl. Math. {\bf 5} (2004) 102.

\bibitem {GD04}
P. D. Gr\"unwald and A. P. Dawid, {\sl
Game Theory, Maximum Entropy, Minimum Discrepancy And Robust Bayesian Decision Theory,}
Ann. Stat. {\bf 32} (2004) 1367--1433.

\bibitem {NJ08}
 J. Naudts, {\sl Generalised exponential families and associated entropy functions,}
Entropy {\bf 10} (2008) 131--149.

\bibitem {NJ09}
 J. Naudts, {\sl The q-exponential family in statistical physics,}
Cent. Eur. J. Phys. {\bf 7} (2009) 405--413.

\bibitem {OA09}
A. Ohara, {\sl
Geometric study for the Legendre duality of generalized entropies
and its application to the porous medium equation,}
Eur. Phys. J. {\bf B70} (2009) 15--28.

\bibitem {OW09}
A. Ohara and T. Wada, {\sl 
Information geometry of q-Gaussian densities and
behaviors of solutions to related diffusion equations,}
J. Phys. {\bf A43} (2010) 035002.

\bibitem {NJ10}
J. Naudts, {\sl The q-exponential family in statistical physics,}
Proceedings of Kyoto RIMS workshop:
"Mathematical Aspects of Generalized Entropies and their Applications",
ed. H. Suyari, A. Ohara, T. Wada, J. Phys.: Conf. Series {\bf 201} (2010) 012003.

\bibitem {NJ11}
J. Naudts, {\sl Generalised Thermostatistics} (Springer Verlag, 2011).

\bibitem {AO11}
S. Amari and A. Ohara, {\sl
Geometry of q-Exponential Family of Probability Distributions,}
Entropy {\bf 13} (2011) 1170--1185.

\bibitem {TC88} C. Tsallis, {\sl Possible Generalization of
Boltzmann-Gibbs Statistics,}
J. Stat. Phys. {\bf 52} (1988) 479--487.

\bibitem {TC10} C. Tsallis, {\sl Introduction to nonextensive statistical mechanics}
(Springer Verlag, 2009).

\bibitem {AS85} S. Amari, {\sl Differential-geometrical methods in statistics,}
Lecture Notes in Statistics {\bf 28} (1985).

\bibitem {AN00}
S. Amari and H. Nagaoka, {\sl Methods of Information Geometry,} Translations of Mathematical Monographs
(Oxford University Press, Oxford, UK, 2000).

\bibitem {TF09}
F. Tops\o e, {\sl
Game Theoretical Optimization inspired by Information Theory,}
J. Global Optim. {\bf 43} (2009) 553--564.

\bibitem {TF11}
F. Tops\o e, {\sl Elements of the cognitive universe,}
\url{http://www.math.ku.dk/~topsoe/isit2011.pdf} (2011).

\bibitem{CI91}
I. Csisz\'ar, {\sl
Why least squares and maximal entropy? An axiomatic approach to inference
for linear inverse problems,}
Ann. Stat. {\bf 19} (1991) 2032--2066.

\bibitem {STD08}
T. D. Sears, {\sl Generalized Maximum Entropy, Convexity, and Machine Learning.}
PhD thesis, Australian National University (2008).

\bibitem {DV10}
Nan Ding and S. V. N. Vishwanathan, {\sl $t$-Logistic regression,}
Adv. Neural Inf. Proc. Systems (2010) \url{http://books.nips.cc/nips23.html}.

% \bibitem {DV10b}
% Nan Ding, S.V.N. Vishwanathan and Yuan Qi,
% {\sl t-Divergence based approximate inference,}
% Adv. Neural Inf. Proc. Systems (2010) \url{http://books.nips.cc/nips23.html}.

\bibitem{BLM67}
L.M. Bregman, {\sl The relaxation method to find the common point of convex sets
and its applications to the solution of problems in convex programming,}
USSR Computational Mathematics and Mathematical Physics {\bf 7} (1967) 200--217.

\bibitem {JE57} E. Jaynes, {\sl Information theory and statistical mechanics,}
Phys. Rev. {\bf 106} (1957) 620--630.

\bibitem {RD69} D. Ruelle, {\sl Statistical mechanics, Rigorous results.}
(W.A. Benjamin, Inc., New York, 1969).

\bibitem {PD07}
D. Petz, {\sl Bregman divergence as relative operator entropy,}
Acta Math. Hungar. {\bf 116} (2007) 127--131.



\end{thebibliography}
\end{document}